# Enlarging the scope: grasping brain complexity

Emmanuelle Tognoli[1]*, J. A. Scott Kelso[1,2]
[1] The Human Brain and Behavior Laboratory, Center for Complex Systems and Brain Sciences, Florida Atlantic University, Boca Raton, FL, USA
[2] Intelligent System Research Centre, University of Ulster, Derry, N. Ireland

**Correspondence:**
Emmanuelle Tognoli
Center for Complex Systems and Brain Sciences
Florida Atlantic University
777 Glades Road
Boca Raton, FL-33431, USA
tognoli@ccs.fau.edu

**Abstract:**
To further advance our understanding of the brain, new concepts and theories are needed. In particular, the ability of the brain to create information flows must be reconciled with its propensity for synchronization and mass action. The framework of Coordination Dynamics and the theory of metastability are presented as a starting point to study the interplay of integrative and segregative tendencies that are expressed in space and time during the normal course of brain function. Some recent shifts in perspective are emphasized, that may ultimately lead to a better understanding of brain complexity.

**Theories of the brain: a concise history**
How does the brain work? This nagging question is an habitué from the top ten lists of enduring problems in Science's grand challenges. Grasp this paradox: how is one human brain –a chef d'oeuvre of complexity honed by Nature– ever to reach such a feast as to understand itself? Where one brain may fail at this notorious philosophical riddle, maybe a strong and diversely-skilled army of brains may come closer. Understanding of the local principles at play has emerged due to the combined efforts of many scientists: neurons talk to their partners by teasing them with charged particles of either excitatory or inhibitory effect, as Nobel laureate Sir John Eccles demonstrated [1]. Targeted release of ions was later shown at sites that seem designed for the exchange of information: typically the axonal termination of the emitting neuron facing the dendrites of a receiving partner [2]. Many of those two-some neural interlocutors build into a reticulum with remarkable emergent properties. A booming network science followed, generalizing microscale principles on a large-scale. David Rumelhart and James McClelland and many others pursued this connectionist endeavor [3,4]. Putting function first, they aimed to model specific aspects of human cognition and behavior such as visual perception or language. Others, such as Olaf Sporns devoted much effort to the "neurobiological" fidelity of their inquiries, conceiving behavior as emergent phenomenon from the appropriate connectional design [5], which they probed either with theoretical connectivity models where brain complexity is carefully thought of [6]; or complementarily with empirically-derived models that borrow their connectional blueprints from images of "real" brains [7].

**Neuronal relays and the propagation of information**
The principle of synaptic transmission proved to be picture-perfect for a theory of communication, boosted by the influential work published in 1948 by Claude Shannon [8]. Transfer of information became a principal tenet of brain function, and theories went so far as to conceive of "centers" as final destinations for information to be communicated (the concept has now retreated, although it remains perniciously present in neuroscientists' conceptions of brain hierarchies; an alternate view is that it is the journey, but not an elusive final destination, that really matters). This theory of information processing in the brain raises a question that may not have received enough attention: can we readily transpose findings from the smallest synaptic level –findings that speak of only a pair of neurons-- to larger spatial scales such as neural areas or the whole brain? In all justice, countless emergent phenomena were discovered through this extrapolation, both in empirical and theoretical work. But it remains an uneasy feeling that so much of Brain Science is built upon the foundation of a pair of neurons, outside the context of their networks, and with two open-ended areas of darkness at either of their extremities that must be thought of as the entire remainder of the organism's brain (and body).

**Collective power of neuronal synchrony**
We will come back to information transmission later, but let us now explore the matter of spatial scales. As humans tend to agree, increased size makes up for smarter brains (disclosure: both authors are human), and those bigger brains have room to organize themselves at multiple levels, coalescing into functional ensembles at several steps along the way up from neurons to functional areas and to the entire brain [9,10,11]. At larger and more integrated levels of description, other ordering phenomena were discovered that brain scientists conceive in terms of information exchange as well. In the late nineteen eighties, two groups of scientists, one with Reinhard Eckhorn [12] and another with Charles Gray and Wolf Singer [13], discovered that perceptual integration (or *Gestalt*) elicited transiently synchronous action potentials amongst neurons that had shared-stakes in the sensory object being viewed. Those neurons dealt with separate parts of the visual field, and they generally disagreed on when to elicit their action potentials in the regular course of their participation in visual function. Somehow however, through the complex labyrinth of the visual cortex and despite the fact that some finite amount of time was required to get from any one to any other of them (delays and frustrations manifested in their usual asynchrony), they managed to coincide when they responded to the same object. What we knew from those neurons is that they "responded" strongly to orientation, fragments of contours with sharp luminance gradients. Their synchrony it seems, was a trace of their joint participation in the construction of something bigger (the object) than what each of them was about (pieces of contour). These discoveries resonated with earlier theorizing regarding the collective behavior of neurons such as Donald Hebb's cell assemblies [14] or Walter Freeman's mass action [15]. The findings by Eckhorn, Singer and Gray launched a relentless quest for synchrony in all parts of the brain and for numerous functions [16], and took the form of several variants (the most basic being coincidence of action potentials and phase-locking of neural oscillations).

**Irreconcilables**
Theories and dedicated experimental paradigms were built upon both discoveries of synaptic transmission and neural synchronization. And from each side, supporting evidence abounded. In spite of their prominence and ubiquity though, the theories carefully avoided confrontation with each other, remaining mostly in the separate territories of distinct research groups. One may note already some difficulties in reconciling them. Let us follow the two extreme views: perfect synchronization and perfect transfer. If all neurons were completely synchronized, they would remain in a changeless state of simultaneity. It is unclear how this system could have flows of information from one place to another. On the other end, if each neuron relayed information in a strict sense, the system would lack basic simultaneity through which synchronous phenomena could emerge. In their radical form it seems, the theories of information exchange *qua* synaptic transfer or neural synchrony are mutually exclusive.

**Can we find directions in the brain?**
The tension is also visible in some empirical facts. Although directed flows of information in the Shannonian spirit do most certainly occur in neural networks, it is indeed quite challenging to track information otherwise than in local or statistical sense (by tracking, we mean to follow the path of information on a brain map as one would follow any object in motion on a symbolic representation of its spatial domain – see figure 1). The brain network after all, is a web, as Francisco Varela emphasized [17], and one gets quickly lost with all the branchings, loops and loops within loops [18,19]; structural features that "distribute" information (albeit unlike a postmaster distributes mail). So it seems that transmission principles do not scale well upward from simple "channels" of synaptic interactions to the larger and more complex web of evolved brains. Thus, it is without surprise that the brain betrays an essential communicational etiquette: its parts do not behave in a sequential - one-talks-at-a-time- manner (as opposed to the humoristic illustration of Figure 1). It is also overwhelmingly clear that "inputs" from the environment do not enter a silent system. Brain parts constantly exchange information about their current and past affairs, and what comes in at a given time works more as a "perturbation" to an already established ballet, an event that weaves itself within a broader scheme of coordinated brain behavior rather than the sole commander of all things present. All of these nuances differentiate the brain from a channel in which information is transferred from sender to receiver. This situation creates mounting complications. The quest for directional flows in the brain has proved difficult both conceptually and methodologically, yet, it has not deterred efforts toward understanding. Mathematical and empirical studies aimed at resolving these questions are an active area pursued by many, including our own colleagues [21,22].

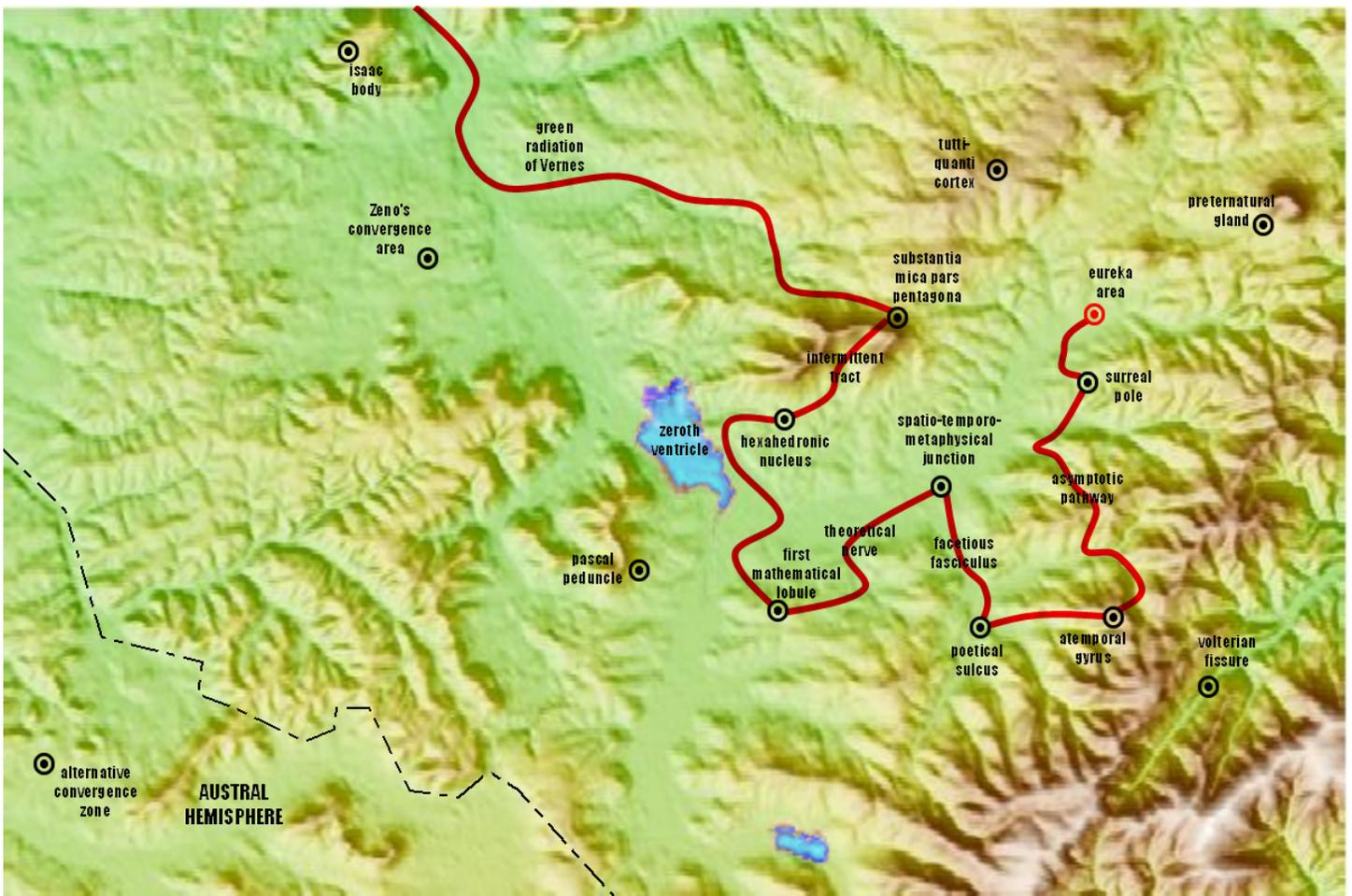

*Figure 1: a teasing figure aimed at marking difference between directionality in well-formed Shannonian systems (as in this imaginary brain map) and in complex systems such as the brain. In the latter, direction is less intuitive past immediate spatial and temporal neighborhoods, and it can reverse across spatial scales of observation [20].The question is highly relevant though, when one is concerned with where and how to effect changes in the system.*

**Brain organization: synchronization or coordination?**
The second concept, synchrony, also bears its share of ambiguities. The firm ground on which we stand is that the timing of neural activity is not left to hazard (as if parts of the brain behaved independently, and were totally oblivious of what the others were doing). "*When"* one brain part behaves influences *when* others do. And like social creatures, neurons also use the power of their numbers to increase their impact, creating collective structures that speak from a common voice. A generic name for such behavior is "coordination" [23]. Synchrony is a narrower concept, one of several ways for a system to coordinate itself. Though synchrony has multiple meanings (and though its study uses a variety of tools across the board), it is easy to conceive and to model, perhaps explaining its systematic resort. To be rigorous however, synchrony requires two important and inter-related characteristics: first, that the underlying temporal order in which the system is embedded be frequency-locked, and second, that attractors have emerged in the system's coordination dynamics (attractors are mathematical structures that entrap the system's coordination dynamics into persistent, -hard to break- states). How to examine if there are attractors in the system from empirical data? We do know how to go from theoretical descriptions of dynamics with- and sans-attractors to their phenomenology (using models to create data at will), but we are not very successful at taking the return path: demonstrating attractors or lack thereof from looking at data, at least for complicated systems like the brain. The other criterion, frequency-locking, is a little bit easier. And what the data say is that brain parts exhibit tendencies toward frequency-locking without going all the way to being perfectly identical. So it seems, we could be dealing with synchronous tendencies rather than synchrony [24]. The difference may seem subtle to some (and some might be tempted to brush it aside as noise or measurement uncertainty), but mathematically, it is enormous: it speaks of two entirely different species of dynamical systems, as said before, one with and the other without attractors. We are turning the spotlight to this distinction because we believe it to be paramount for progress in understanding the brain.

**At the crossroads of propagation and synchronization**
We hope that the previous exposé motivated the thought that neural networks neither operate on perfect synchrony nor on strict transfer, which is good news as each prevents expression of important features of the other. What then is the link between them? Some attempts at studying synchrony and transfer in a common formalism have emerged, exceptions to their usual avoidance of one another. One is the quest for quantifying directional coupling as discussed above. Another attempt is functional connectivity, a daring concept that Karl Friston created on his way to developing theoretical and computational tools for the analysis of functional images of the brain [25]. Connectivity deals with ways for information to go from one place to another. Ideally, we would be able to measure the connection (the "traffic" between two sites) independently from the state of those sites where said traffic imparts effect (as one would measure how many cars travel on the road between two cities). If independently measured, large scale connectivity and local activity would be amenable to reveal their effect on each other. Since we do not have adequate tools to measure the flow of information in living fiber tracts at large though, connectivity is not measured directly; rather it is inferred from the way brain components behave. Interaction, it is postulated, has to leave detectable traces in the behavior of its participants. Of course, contemporaneous theories have shaped the lens through which scientists have tried to see this influence. To make things practical, the assumption was often made that regions exchanging information must be correlated or synchronized (connectivity→correlation). Flipping things around for the operational goal of quantifying the unquantifiable information flow, "how much regions were correlated" became the proxy for how much they exchanged information (correlation→connectivity). But with only this concept of synchronization under the scope, we may see a mere fraction of the brain at work, the tip of the iceberg. What if most coordinated behaviors in the brain do not fall under our definition of correlation or synchrony? Depending on the methods used, that would mean for instance brain regions that are coordinated yet not temporally coincident; or assemblies in which self-organization favors a fluid coordination regime sans-attractor (such as metastability, to be discussed below) over rigid states of phase-locking. Can we see dynamics in which no absolute "order" emerges in space (synchronization) or in time (transfer), and still make sense of it as a means for the brain to function? Those are the dark and uncharted areas in the spatiotemporal organization of complex systems – those for which we sorely lack concepts and methods (Figure 2).

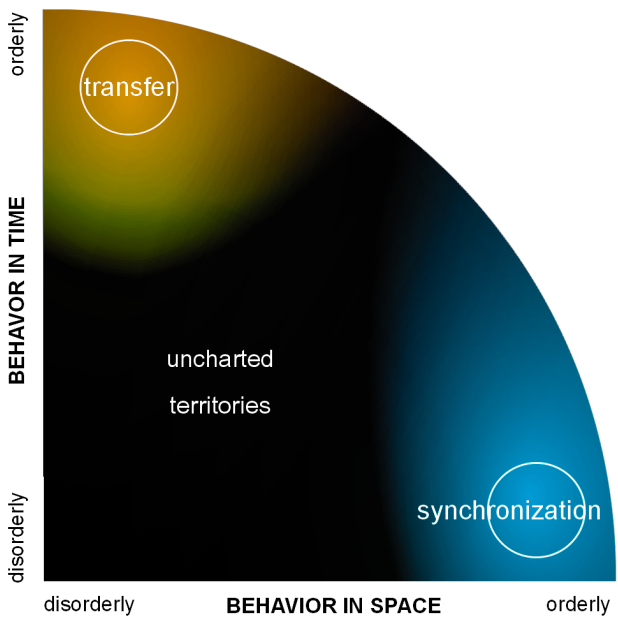

*Figure 2: a graph of spatiotemporal organization. The horizontal axis quantifies the degree of order in space; order in time is on the vertical axis. Examples of orderly phenomena are propagation and synchronization. There are numerous hints that the brain is more efficient with less than complete order in space and time. The dark area represents this region with incomplete spatial and temporal order, for which concepts and tools are sorely missing.*

**Enlarging the scope: metastability**
The set of questions above resonates with a recent shift in perspective on brain function, from a primary focus on neural synchronization to the broader—and deeper-- problem of dynamic coordination. This shift was salient in the editorial introduction to a special issue of Nature Review Neuroscience in February 2010, where the word "coordination" occurred 6 times in a short text of 250 words [26]. And this is a concept that is growing [11,16,20,23-34]. Coordination includes synchronization as one possible collective behavior, but it also considers many other ways for components of the brain to interact. In particular, under certain conditions partially synchronized behaviors arise. In them, the parts exhibit simultaneous tendencies to temporarily couple and to segregate as independent entities. Such "metastable" regimes, we and others have shown, constitute a recipe for complexity [23,11,20,24,28-34]. Why is metastable coordination dynamics of high interest to understanding how the brain works? First, it typically arises when the parts are no longer perfect clones of one another (e.g. as in computational models built from collections of identical neurons). When symmetry is broken and interacting parts are recognized in the diversity of their intrinsic behavior, a more realistic picture of brain function emerges (indeed a trend toward studying more diverse associations in the brain may explain the shift in perspective, for instance the interactions between neurons and astrocytes). Second, incomplete synchronization is more adaptive than pure forms. A fundamental nonlinearity in brain self-organization exists. Too much autonomy (parts of the brain hardly ever affected by what others are doing) prevents emergence, integration and mass action [35]. Yet, too much integration (for instance the whole brain engaged in a giant common behavior, [36]) is inadequate too, because the respective parts can no longer do what they are supposed to do in contributing to collective behavior. The parts then have no choice but to behave exactly like each other and the richness of their individual dispositions is lost to the ensemble. It is enough to note –as many have--that excess synchronization is pathological in the brain, for instance in epilepsy or Parkinson's disease [35,37-38]. As a result, the ideal place for a brain to exhibit a rich set of meaningful behaviors is in-between integration and segregation. This is where the "incomplete" synchronization tendencies - or metastable coordination comes into play [20,23-25,29,31,39]. Elsewhere, we have also speculated on the tremendous functional advantages that metastability would confer to a system, including speed, flexibility and resilience [32].

**Creating a new conceptual framework**
The gap between our current understanding of the brain and the miracles of our mental life and behavioral achievements (for example, consciousness and capacity for invention) remains abysmal. Looking through the history of science, several paradigms of brain function flourished and then dried up following the ebbs and flows of scientific metaphors. The ultimate model, the one that allows to forecast all matters of brain action and to design an artificial counterpart of multiple functional prowesses, remains out of sight. Two lines of thinking have been much explored in recent times: information transfer and synchronization. Their success owes much to the fact that they are special cases and open to quantification. When examined together though, they reveal some incompatibilities that seem to require a relaxation of both principles: less stringent temporal order and less complete spatial order. To advance our understanding of the brain, Neuroscience must open up avenues to study functional behavior in a broader sense. We face two alternatives: to leave it all within the current framework, with the approximate truth derived from current theories (the brain "sort of transfers information", and it operates with "near synchrony"), or to face the issue head on with a different theoretical mindset. In the latter case, a new phenomenology is up for grasp. It will be difficult to conceptualize, and even more so to observe, since it points toward a void in understanding. Modeling approaches can lead the way, by informing which observables we can expect to encounter in the coordinating living brain. And tools will have to be revised or built to adapt to this new world, tools that will say for instance, when "more synchrony" is "too much synchrony" (astonishingly, this simple question is not built into our current enquiries, despite obvious evidence of the ills of excess synchrony). We note that Brain Science is reaching a turning point that may make this renewal possible: it shows many signs of its readiness to enlarge the scope on brain function, not least of which is a recent outburst of interest in segregation phenomena [40-42]. A new paradigm would help to integrate principles that seem contradictory in their radical form: transfer and synchronization, as well as integration and segregation. Those pairs of concepts are reconciled under the dynamical regime of metastability [20,43].

**Acknowledgements**
This work was supported by NIMH (MH080838), NSF (BCS0826897), the US ONR (N00014-09-1-0527), the Davimos Family Endowment for Excellence in Science and the Chaire d'excellence Pierre de Fermat.